\documentclass[final,3p,fleqn,a4paper]{elsarticle}

\usepackage{amsfonts,amsmath,amssymb}				
\usepackage[english]{babel}					
\addto\captionsenglish{}
\usepackage{graphicx}						
\usepackage{wrapfig}
\usepackage{multicol}						%
\usepackage{color}
\usepackage{listings}						
\usepackage{anysize} 						
\usepackage{latexsym} 
\usepackage{subfig}
\usepackage{multicol}						%
\usepackage{multirow}	
\usepackage{breqn}
\usepackage{gensymb}%
\usepackage[section]{placeins}
\usepackage{textcomp}
\usepackage{gensymb}

\newcommand{\reffig}[1]{Figure~\ref{#1}}

\newcommand{\reftab}[1]{Table~\ref{#1}}

\newcommand{\refsec}[1]{Section~\ref{#1}}

\begin{document}

\begin{frontmatter}
\title{Crack arrest through branching at curved weak interfaces: \\an  experimental and numerical study}

\author[US]{M.T.~Aranda}
\ead{maranda2@us.es}

\author[US]{I.G. Garc\'{i}a}
\ead{israelgarcia@us.es}

\author[US]{J.~Reinoso}
\ead{jreinoso@us.es}

\author[US]{V.~Manti\v{c}}
\ead{mantic@us.es}

\author[M.Paggi]{M. Paggi}
\ead{marco.paggi@imtlucca.it}

\address[US]{Grupo de Elasticidad y Resistencia de Materiales, Escuela T\'{e}cnica Superior de Ingenier\'{i}a, Universidad de Sevilla \\ Camino de los Descubrimienos s/n, 41092 Sevilla, Spain}
\address[M.Paggi]{IMT School for Advanced Studies Lucca, Piazza San Francesco 19, 55100, Lucca, Italy}

\journal{Theoretical \& Applied Fracture Mechanics}

\begin{abstract}
The phenomenon of arrest of an unstably-growing crack due to a curved weak interface is  investigated. The weak interface can produce the deviation of the crack path, trapping the crack at the interface, leading to stable crack growth for certain interface geometries. This idea could be used as a technical solution for a new type of crack arrester, with a negligible impact on  the global stiffness, strength and weight of the structure. In order to exploit this concept, an experimental campaign based on photo-elasticity and digital image correlation is carried out, showing the capability of curved weak interfaces to arrest cracks. The experiment is repeated for several geometrical configurations through the modification of the interface curvature radii.
The phenomenon of crack deviation and subsequent arrest at the interface is also investigated with the assistance of a computational model based on the finite element method. The computational predictions provide the rationale for the interpretation of the experimental observations, and distinguish between the different behaviour of concave and convex interfaces. Consequently, as is shown in the present study, the curved interface concept fosters new routes for the attainment of structures with enhanced fracture resistance capacities, which are of paramount importance for materials and components used in 
extreme conditions.

\noindent \emph{The final version of this manuscript has been published in Theoretical and Applied Fracture Mechanics
Volume 105, February 2020, 102389}.
\end{abstract}


\end{frontmatter}

\section{Introduction} \label{sec-CurInt1-Intro}

The damage-tolerant approach has become a recurrent standard for the design of structures in aerospace and ship-building industries, among others, see e.g.~\cite{braga2014}. The objective of this methodology is the specific design of structures tolerant to the presence of flaws, i.e.~where the presence of cracks does not imply necessarily the  catastrophic failure of the structure upon their propagation.
Following this idea, many attempts have been made in materials science in order to identify optimal material microstructures to arrest cracks, also inspired by nature, see e.g.~\cite{gao1,gao2,gao3,PW2012}.    
However, the application of this approach generally requires: ($i$) advanced inspection techniques in order to detect damage events prior they  reach  the corresponding  critical sizes, and ($ii$) the design of structural elements that enable either preventing or slowing-down  the progress of crack evolution and the related structural softening. Within this context, the so-called crack arresters can be understood as structural elements that are designed  for  this purpose. In the last years, such a concept  has been prolifically employed in different practical applications in engineering,  see e.g.~\citep{lena2008,somerville1982}. In materials science, the role of structural arresters can be played by material interfaces, suitably arranged across the scales in order to prevent the catastrophic propagation of flaws. For instance, \cite{PW2012} have shown that a two-scale arrangement of material interfaces can be profitable to stop the propagation of cracks from the microscale to the mesoscale in polycrystalline materials used for extreme structural applications such as drilling. 

With the aim of exploiting the role of material interfaces, we consider here weak interfaces, which are material discontinuities favorable to debond and deviate cracks  by investigating their curvature. It is in fact well-known from the fundamental studies by \cite{he1989,hutchinson1992} that the competition between crack branching and penetration when a crack impinges into an interface can lead to different scenarios, such as crack penetration, single deflection along the interface, and also double deflection, depending on the elastic mismatch between the two joined materials and the toughness of the interface as compared to that of the bulk. Those results, however, were restricted to flat interfaces, while interfaces with curvatures could also be explored and manufactured. The sustained interest emerges from characteristic material disposals in many engineering systems, ranging from composite materials, presence of bonding lines, curved fibers, among many other scenarios. Therefore, the profound understanding of crack propagation at interfaces can be conceived as a key aspect that can actively contribute to the design of new generation of composite materials and structures with enhanced fracture resistance properties. 

Reference studies in this context can be traced back to \cite{cook1964}, who established a strength-based criterion to determine the potential crack path through the stress field estimations near the crack tip recalling the \cite{Inglis1913} procedure. Posterior investigations advocated the use of linear elastic fracture mechanics (LEFM), which exploited an energetic approach to predict crack penetration vs.~crack deflection scenarios in multi-material systems. LEFM-based procedures generally require the computation of the ERRs corresponding to a perturbation of the linear elastic solution by introducing infinitesimal kinks associated with crack penetration or crack deflection, see \cite{hutchinson1992,leguillon2000,martin2001,martin2008}. These contributions have been of notable interest to many researchers in order to interpret many experimental data, pinpointing the necessity of developing reliable numerical and theoretical procedures in order to achieve a more thorough comprehension of such phenomena. 

Subsequently, from a computational standpoint, notable works have been conducted through the use of the so-called cohesive zone model  (CZM) to predict penetration vs. deflection scenarios. The fundamentals of the CZM were originally proposed in \cite{barenblatt1962}  and  \cite{dugdale1960}, who considered a combined strength-and-energy approach to trigger fracture events in solids. Within this context, \cite{Parmigiani2006} analyzed these phenomena using the CZM in media with the presence of weak interfaces, showing that the fracture toughness and the strength of the bulk and of the interface play a significant role in order to determine whether the crack will deflect along the prescribed interface or, conversely, it will propagate into the adjacent body. Specifically,  it has  been addressed that higher interface strength promotes crack penetration, whereas weak interfaces induce crack deflection along existing interfaces, examining a situation where the interface is perpendicular to the traveling mother crack.  However, one of the main limitations of the CZM relies on the fact that: ($i$) it implies the pre-definition of the potential crack paths prior running the simulations, in its implicit version \cite{PW12}, or ($ii$) it requires the incorporation of complex crack tracking algorithms using the explicit CZM, confining the potential crack paths to the boundaries of the corresponding discretization  \cite{Xu1994} (mostly in the spirit of the finite element method, FEM). In contrast to the previous computational methodology, the advent of the phase field method (PFM) of fracture \cite{Bourdin2008,Miehe2010}, which is based on the regularization of the variational formulation of brittle fracture due to \cite{Francfort1998},  does offer a potent modeling alternative with strong potential for simulating very complex fracture phenomena.  With respect to  heterogeneous media,  \cite{paggi2017revisiting}  devised the consistent coupling of the PFM with  the CZM (PF-CZM) which is especially oriented to the simulation of fracture events with the presence of weak interfaces. The predictive capabilities of the seminal approach developed in \cite{paggi2017revisiting} has been shown through its application in layered ceramics \cite{CAROLLO20182994}, laminates \cite{carollo20173d}, polycrystalline materials \cite{paggi2018fracture}, and micro-mechanics of composites \cite{garciapaggimantic2014}, retrieving also fundamental results from LEFM \cite{hutchinson1992} concerning bi-materials systems.  

A plausible alternative methodology to the previous procedures is the so-called Coupled Criterion (CC) \cite{leguillon2002} or Finite Fracture Mechanics (FFM), whose rigorous development implied the simultaneous satisfaction of the corresponding energy- and strength-based criteria with excellent accuracy capabilities. CC-based method permits to gain a plausible insight into the physical events under investigation yielding to persuasive explanations with the precise application of both strength and energy criteria. The application of the CC or FFM has successfully applied in many engineering problems \cite{martin2012}, material bonding \cite{weissgraeber2011}, thermo-mechanical studies \cite{leguillon2013-thermal,garciatermico} and more specifically within the context of laminate composite materials \cite{manticgarcia2012,garcia2014-IJSS}, among many others. 

From the previous brief review, it becomes evident that the investigation of the competition between crack deflection and propagation at weak interfaces has been long identified as one of the topics that has attracted a great deal of research in the last years. Comprehensive  experimental  studies have investigated the problem of interfacial crack growth \citep{ROSAKIS19981789,CHALIVENDRA20082385}. Thorough investigation on this specific application complying layered composite materials have been conducted in \cite{ALAM2017116} and by Tippur and co-authors \citep{SUNDARAM2016312}. It is also worth mentioning the development from this researcher group of new experimental techniques to trigger dynamic crack events using the  a novel vision-based method denominated as Digital Gradient Sensing (DGS) which is combined with ultrahigh-speed photography \citep{SUNDARAM2018132}. Nevertheless, most of the existing studies in the related literature have been principally devoted to the analysis of straight interface definitions following either perpendicular \cite{SUNDARAM2016312} or inclined directions \cite{ALAM2017116} with respect to the primary crack. However, at present, situations with the possible presence of curved and/or patterned interfaces, that yield to outstanding fracture resistance properties  \cite{cordisco16,garcia-guzman,BERTOLDI20071,BIGONI20024843}, has received a very scarce attention. Consequently, motivated by these real scenarios, the present study is devoted to evaluate the ability of curved weak interfaces to serve as crack arresters, i.e. carefully investigating crack deflection vs. propagation in specimens with curved interfaces. 

In the sequel, after a description of the experimental setup, a detailed analysis of the results and the corresponding interpretation is carried out. Thus, the manuscript is organized as follows. 
The experimental campaign will be presented in \refsec{sec-CurInt1-ExpCamp}. The results obtained will be analyzed with the assistance of a computational model described in \refsec{sec-CurInt1-CompAna}. The discussion of the results is included in \refsec{sec-CurInt1-Disc}, whereas the main conclusions are summarized in \refsec{sec-CurInt1-Conc}.

\section{Experimental campaign} \label{sec-CurInt1-ExpCamp}

The objective of this section is to analyze the viability of the design of a special geometry of a weak interface to \emph{retard or modify} the crack path, by means of deviating the (mother) crack along such an interface. In addition to deviate the crack, the role played by different interface curvatures onto the overall structural response are carefully investigated. 

In the following, current experiments  will be presented in relation to: the geometry of the specimens, their fabrication procedure and  materials used, and the method and experimental setup. Finally, details regarding  the experimental results for different interface curvatures and positions with respect to the mother crack will be put forward.   

\subsection{Specimens design and fabrication} \label{sec-CurInt1-ExpFab}
\begin{figure}[ht!]
  \centering
     \includegraphics[width=1\columnwidth]{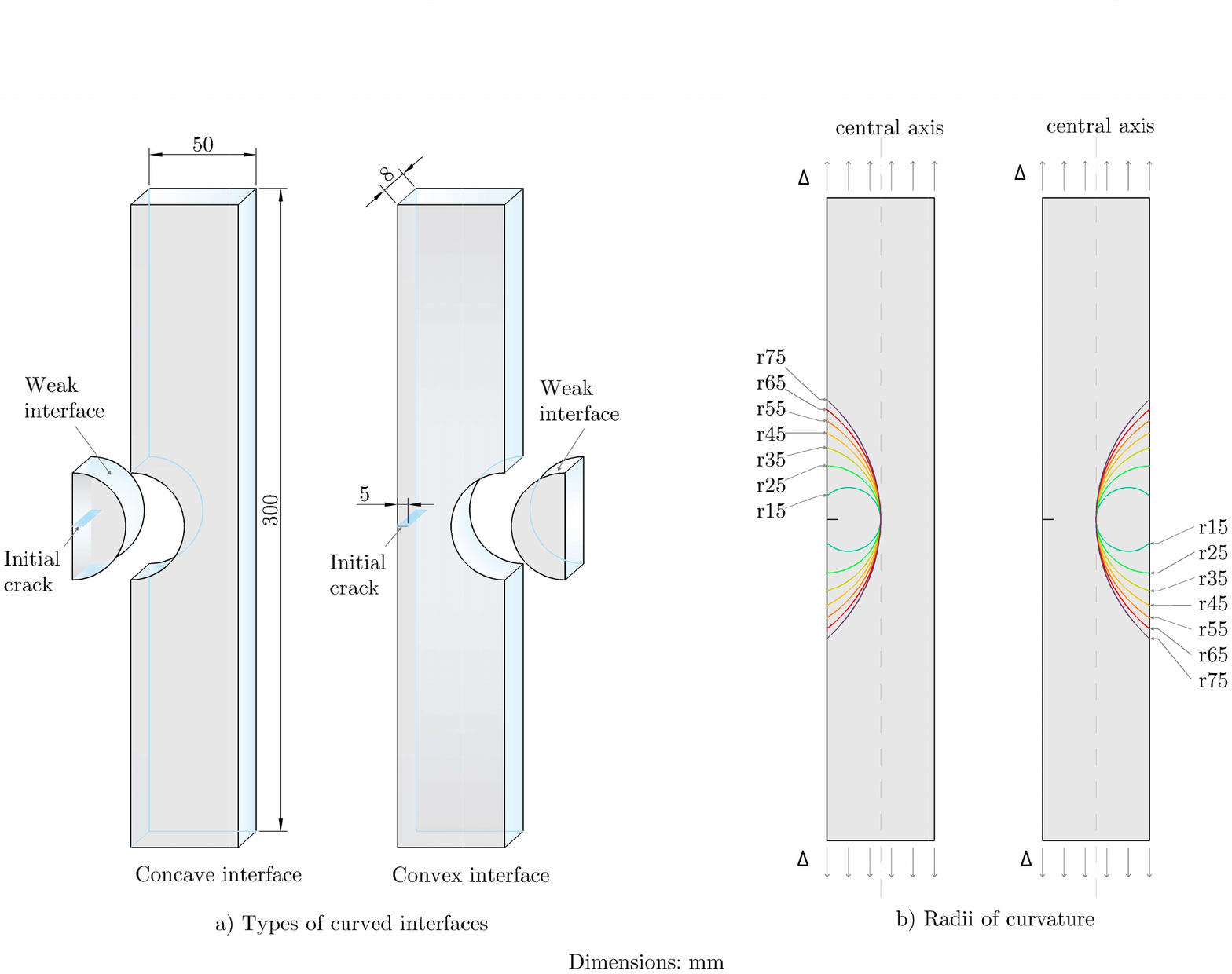}
     \caption{Sketch of the studied specimens geometry. a) Concave and convex weak interfaces. b) Different radii of curvature.}
         \label{fig:esquemaprobeta}
 \end{figure}

Focusing on the study of the effect of the curvature radius of a weak interface on the propagation of a dynamic crack, appropriate  specimens were designed including a novel concept of curved interface,. This interface bonds two parts of a rectangular specimen with dimensions $300\times50\times8$ mm. Two types of specimens with an initial crack and a curved interface were fabricated, namely: ($i$) the so-called \emph{concave curved interfaces}, where the initial mother crack is approaching a concave (non-convex) part of specimen (i.e. approaching an interface point from the side of the osculating circle), and  ($ii$) the so-called \emph{convex curved interfaces}, where the initial mother crack is approaching a convex part of the specimen (i.e. approaching an interface point from the side opposite to the osculating circle), as shown in the sketch in \reffig{fig:esquemaprobeta}a. For each type of interface, different  curvature radii were studied, ranging from 15 mm to 75 mm, with discrete variations of 10 mm each. Such interfaces were located tangent to the central longitudinal axis of the specimen, as shown in \reffig{fig:esquemaprobeta}b. In this way, 14 specimens were patterned, seven of them corresponding to concave interfaces and other seven configurations with convex interfaces.

The material selected for the fabrication of the specimens was polymethyl methacrylate, known as PMMA, because it is a brittle material suitable for photo-elasticity within a certain application range. Moreover, from the practical point of view, it was easy to form bonded interfaces with PMMA using commercial solvents. Note that PMMA was frequently used in previous experimental studies for interface fracture, see \cite{sundaram2016} and the references therein given. Likewise, the interface was made of a transparent, bi-component epoxy resin with a viscosity of $6.000$/$9.000$ cps and a density of $1.10$ gr/cc, a material which could also be used for photo-elasticity \citep{Noselli2010,DALCORSO2008815,MISSERONI201487}.

After several preliminary manufacturing and testing trials to assess the best bonding conditions, the  specimens were manufactured by cutting them with a laser cutter of 150 W, splitting the specimen into two parts. Subsequently, an initial crack (pre-crack) of approximately 5 mm length and oriented perpendicular to the edge was introduced in each specimen  by a razor blade, as recommended in~\cite{ASTMD5045} and already used for other experiments with the same material, see e.g.~\cite{leite2014}. In a next stage, the specimens were prepared to bond the two parts by cleaning the contact surfaces. During this procedure, the initial mother crack was covered to prevent undesired leakage of the adhesive into it.

The two parts of the specimen were bonded together using a special fixture under controlled pressure to attain  and manage the proper adhesive thickness. Then, the specimens were left curing  at room temperature for a period of 6 days, in order to achieve the required bonding strength and avoid residual stresses that would affect photo-elasticity measurements. The excess of adhesive was not removed, precluding the pore formation at the interface. Finally, an adhesive drop of pipette was provided at the intersection line of the interface and the specimen edges, in order to prevent premature failure near such critical zone close to the boundaries. 


\subsection{Experimental setup} \label{sec-CurInt1-ExpSet}

\begin{figure}[ht!]
  \centering
     \includegraphics[width=1\columnwidth]{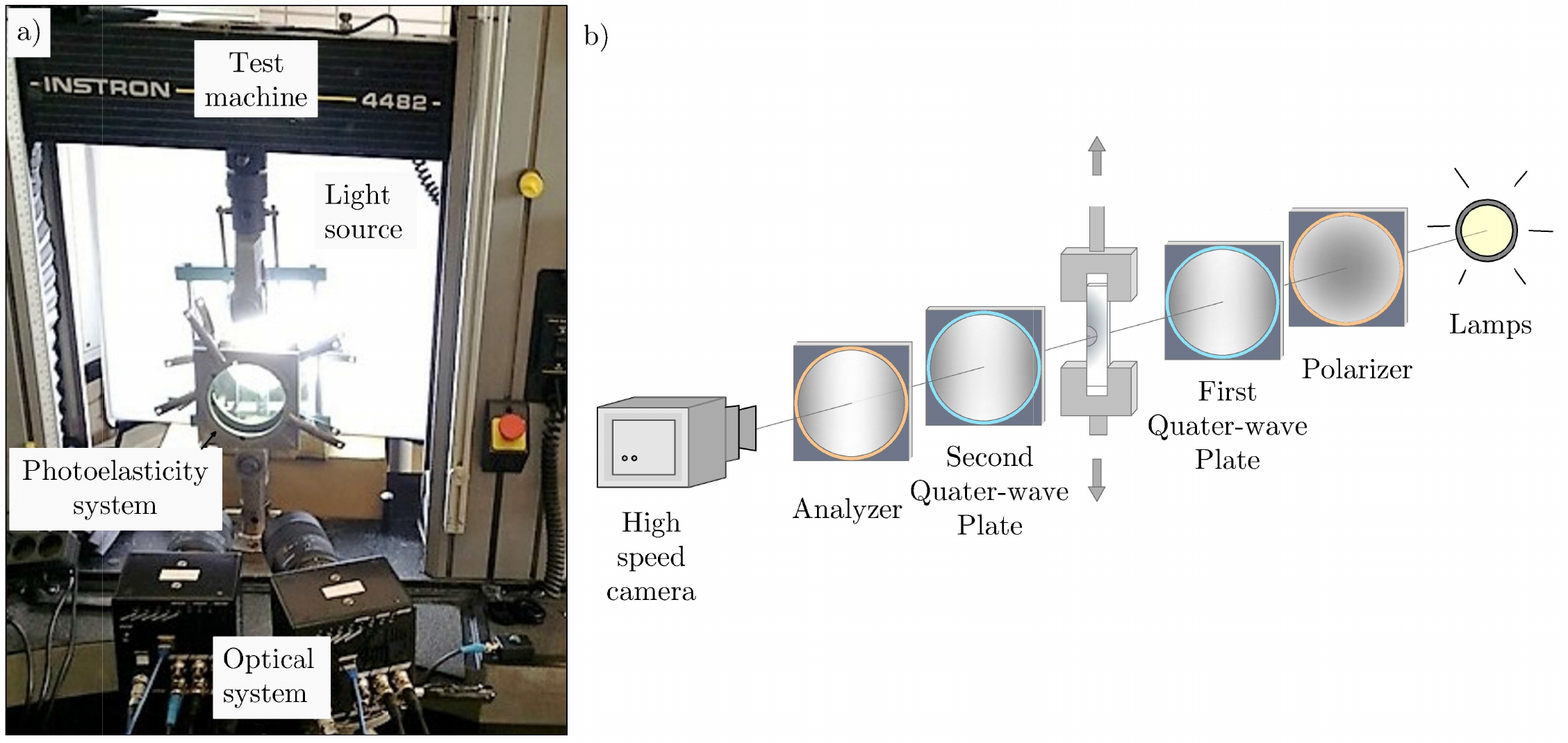}
     \caption{a) Close-up photograph of the experimental setup in this study. b) Schematic of the photo-elasticity setup. }
         \label{fig:equipos}
 \end{figure}

The experimental setup, see \reffig{fig:equipos}, consisted of: ($i$) a mechanical system to apply a tensile   loading on the specimen, ($ii$) an optical system to produce isochromatic fringe patterns to highlight the crack path during propagation, and ($iii$) an imaging system to record the photo-elastic fringe patterns and the specimen during crack growth. 

The mechanical system involved an Instron 4482 universal testing machine, which was used to test all the specimens under monotonic tensile conditions up to failure. The testing machine was equipped with a 100 kN load cell, being the tests conducted under displacement control mode. Due to the longitudinal dimensions of the specimens, i.e 300 mm in length, they 
were fixed to the testing machine using a standard gripping,  with negligible influence of  the clamping area on the fracture process region within the coupon.  The loading application was carried out by the adoption of a uniform applied displacement of  0.5 mm/s during the tests. 

The imaging system was composed of two Fastcam Mini Ax-200 monochrome high speed cameras, which are able to record images at rates up to 0.9 million frames per second (fps). In our experiments, the  frequency  of recording  ranged from 48,000 to 160,000 fps, depending on the area observed and the resolution prescribed for the camera. Two regions of interest were defined in this study: \textit{(i)} The initial mother crack and the subsequent crack path up to reaching the interface. Accordingly, one camera was focused on a strip line of 512 $\times$ 48 pixels region at 160,000 fps to register a detail view of the first propagation. \textit{(ii)} The second region of interest, i.e. the interface location, was recorded using another camera focusing on a 384 $\times$ 288 pixels region to register a view at 48,000 fps. This optical method was chosen for the tests because it allowed the closest observation of the propagation of a fast running crack to be performed. To illuminate the specimen, two strong flash lamps were used, composed of 4 heads of LED lamps of high power and low consumption model QT of GSVitec.



With regard to the  photo-elasticity system, the usage of this equipment has been adopted due to the fact that it permits a rapid and efficient identification of the crack trip along the propagation of the crack within the specimen. It regarded of a polarizer, an analyzer, two quarter-wave plates between which the model under testing was introduced, and a light source, see \reffig{fig:equipos}b. Hence,  the first quarter-wave plate was placed between the polarizer and the specimen, whereas the second quarter-wave plate was located between the specimen and the analyzer. Therefore, as stated above, the  principal aim was to  qualitatively visualize stress field during crack propagation in monochromatic colors in order to track the crack advance and detect premature debonding along the weak interface.


\subsection{Experimental results} \label{sec-CurInt1-ExpRes}
This section presents the results obtained from the experimental study with the aim of obtaining qualitative conclusions and highlighting the observed trends. The first part describes the behaviour  during the tests, differentiating between convex and concave curved interfaces. The second part analyzes several series of fringe pattern pictures corresponding to a crack propagation in the material in the presence of a weak interface. Finally, photographs of fractured specimens are presented, examining the effect of the interface geometry with respect to the initial crack position.

\subsubsection{General response of specimens}
\begin{figure}[ht]
  \centering
     \includegraphics[width=1\columnwidth]{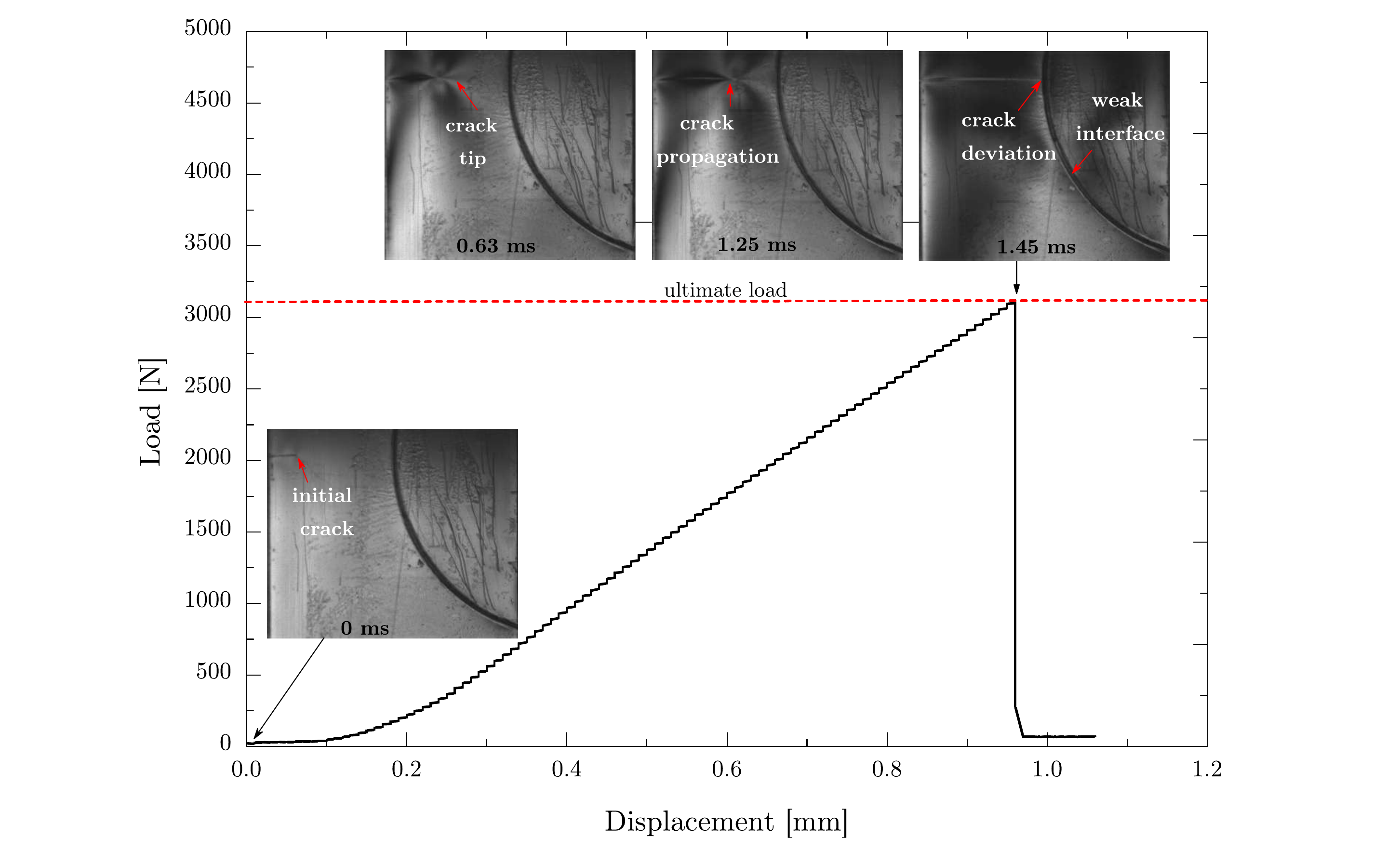}
     \caption{Load-displacement curve of the convex curved interface with a radius of curvature of 35 mm under tensile test. Magnification views of the crack and interface at the zero and the maximum load before catastrophic fracture, 48000 fps, 384 $\times$ 288 mm.}
         \label{fig:convex}
 \end{figure}

\reffig{fig:convex} shows the load-displacement curve for a convex curved interface with 35 mm radius. In this particular load-displacement curve, the ultimate load is $3.1$ kN. For this ultimate loading, a fast catastrophic crack growth can be observed in the three images shown on the top of the figure recorded at 48,000 fps. These images show a general view of crack propagation through the material until its reaches the interface, $1.45$ ms after the initiation of crack propagation. At this moment, the crack is deflected towards the interface, progressing unstably through it and leading to the total failure of the specimen with a catastrophic drop in the load-carrying capacity. 

For all the specimens with a convex interface, regardless their curvature radius, the same brittle response was noticed. Thus, the mother initial crack, that was generated during the manufacturing  process, progressed under mode-I conditions up to reaching the interface. At this stage, this crack was trapped by the interface, and bifurcated into two interface cracks that grew simultaneously along the interface, separating the specimen into two parts. In every test, the crack extent along the interface corresponded to its entire length, so that such interface cracks did not kink out of the interface and, thus, in particular did not penetrate into the other part of the specimen. Moreover, from a qualitative standpoint of the analysis of the load-displacement evolution curve (\reffig{fig:convex}), it is observable that the specimens featured a quasi perfect linear elastic evolution up to an abrupt failure, at which the coupons broke in two parts. 

\begin{figure}[ht!]
  \centering
     \includegraphics[width=1\columnwidth]{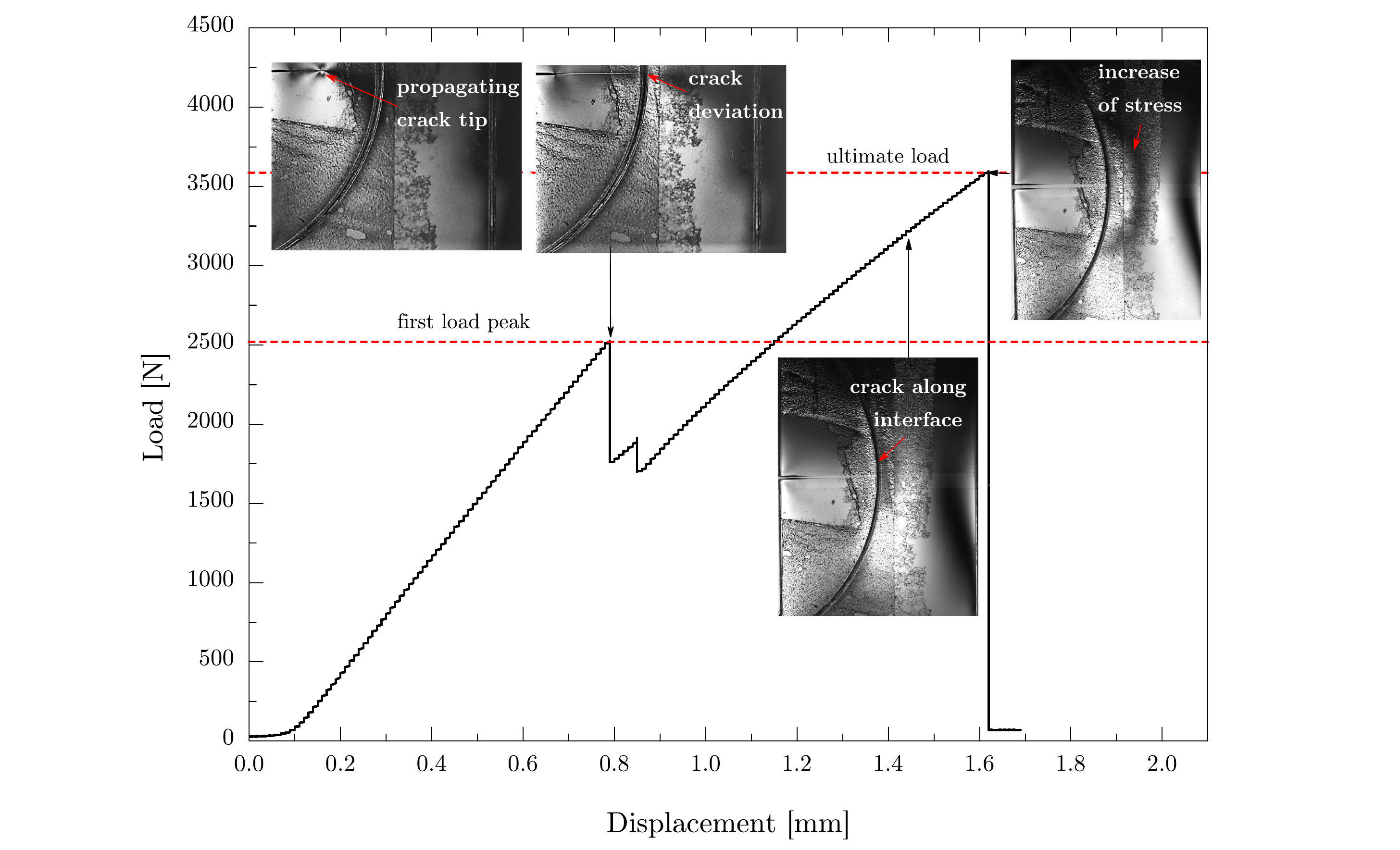}
     \caption{Load-displacement curve of the concave interface with a radius of curvature of 45 mm under tensile test. Magnification views in the first kink load point and at the ultimate load point, 48000 fps, 384 x 288 mm.}
         \label{fig:concave}
 \end{figure}

Interestingly, in the case of concave curved interfaces, the results were drastically different. \reffig{fig:concave} depicts a representative  load-displacement curve for one of the specimens tested with a concave curved interface, displaying three distinct stages of crack propagation. Analyzing this plot, a first linear load-displacement branch can be observed up to the first peak load. Throughout the first stage, the mother crack propagated under mode-I conditions till reaching the interface. After the   crack bifurcation at the interface, the second stage   was characterized by subsequent (approximately symmetric) growth of two interface cracks, one propagating upward and the other downwards, and  featuring a stable growth with an essentially steadily increasing load   with some unstable jumps, as can be observed in \reffig{fig:concave}. In the last third stage, the final failure occurred when one crack emerges from one the the interface crack, kinking out of it towards the other part of the specimen, growing across the bulk  in an unstable manner again. Note that this ultimate loading level was significantly higher than that corresponding to that which produced the first crack propagation. This means that the presence of a concave weak interface significantly increased the load-carrying capacity of the specimen.

At this point, it is worth mentioning that the present results clearly evidenced that the interface geometry, especially its curved shape,  plays  a crucial role in the structural response. Thus, the insertion of a weak interface does not always produce crack penetration into the other part of the body. Therefore, the location and shape of such an interface within a given specimen   and for some loading conditions   may influence the creation of new daughter cracks, as discussed in the following. 

\subsubsection{Analysis of fringe patterns and fractured specimens}
\reffig{fig:concave} shows a set of images corresponding to the first peak load, showing the propagation of the initial crack until it impinges on the weak interface. From this point on, the propagation of the crack along the interface can be tracked. In these images, it can be observed that a part of the interface remains bonded and the region enclosed  by the interface and mother crack flanks is unloaded. Thus, to break the other load-bearing part of specimen, including stress singularities at the interface-crack tips and a stress concentration in its net cross-section at the already broken interface, it is necessary a significant load increase, as discussed previously.

\begin{figure}[ht!]
  \centering
     \includegraphics[width=0.74\columnwidth]{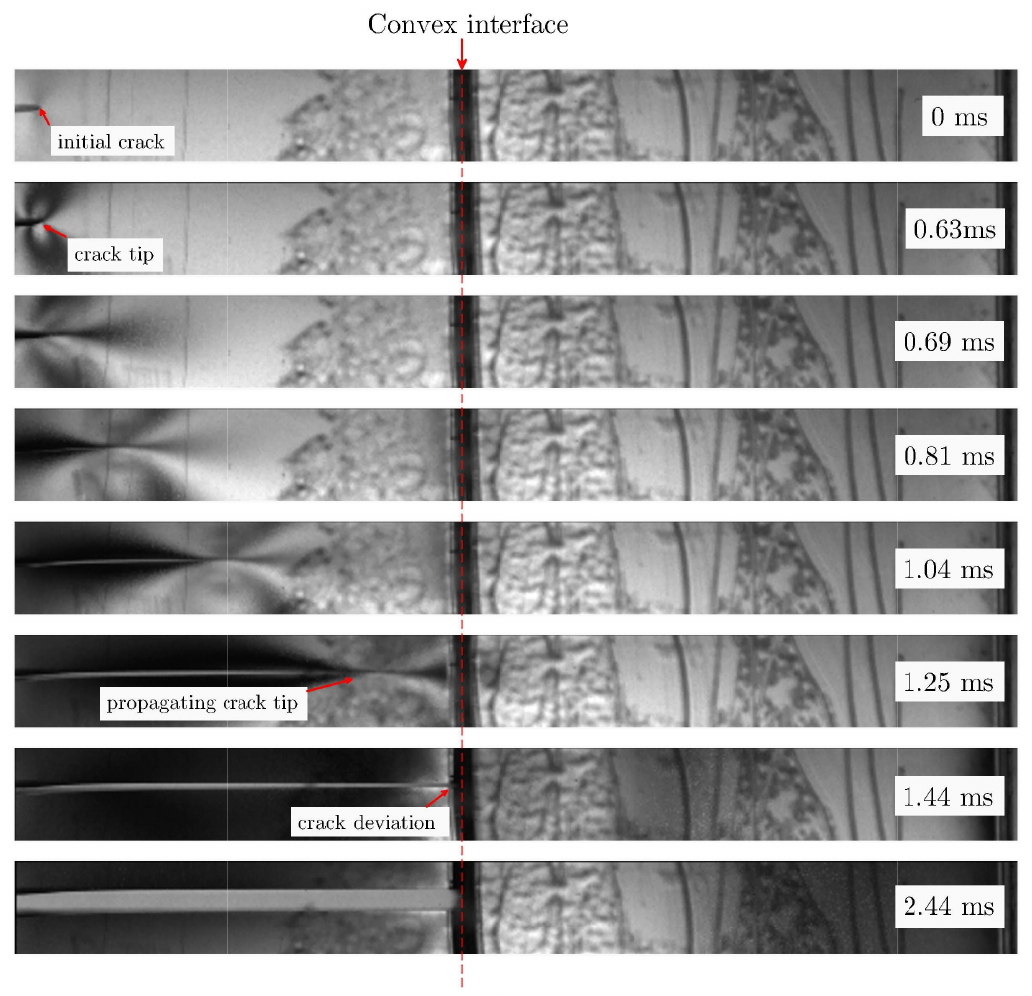}
     \caption{Series of dynamic photoelastic images of crack propagation for a 35 mm radius weak convex interface.}
         \label{fig:detalleconvex}
 \end{figure}

\begin{figure}[ht!]
  \centering
     \includegraphics[width=0.74\columnwidth]{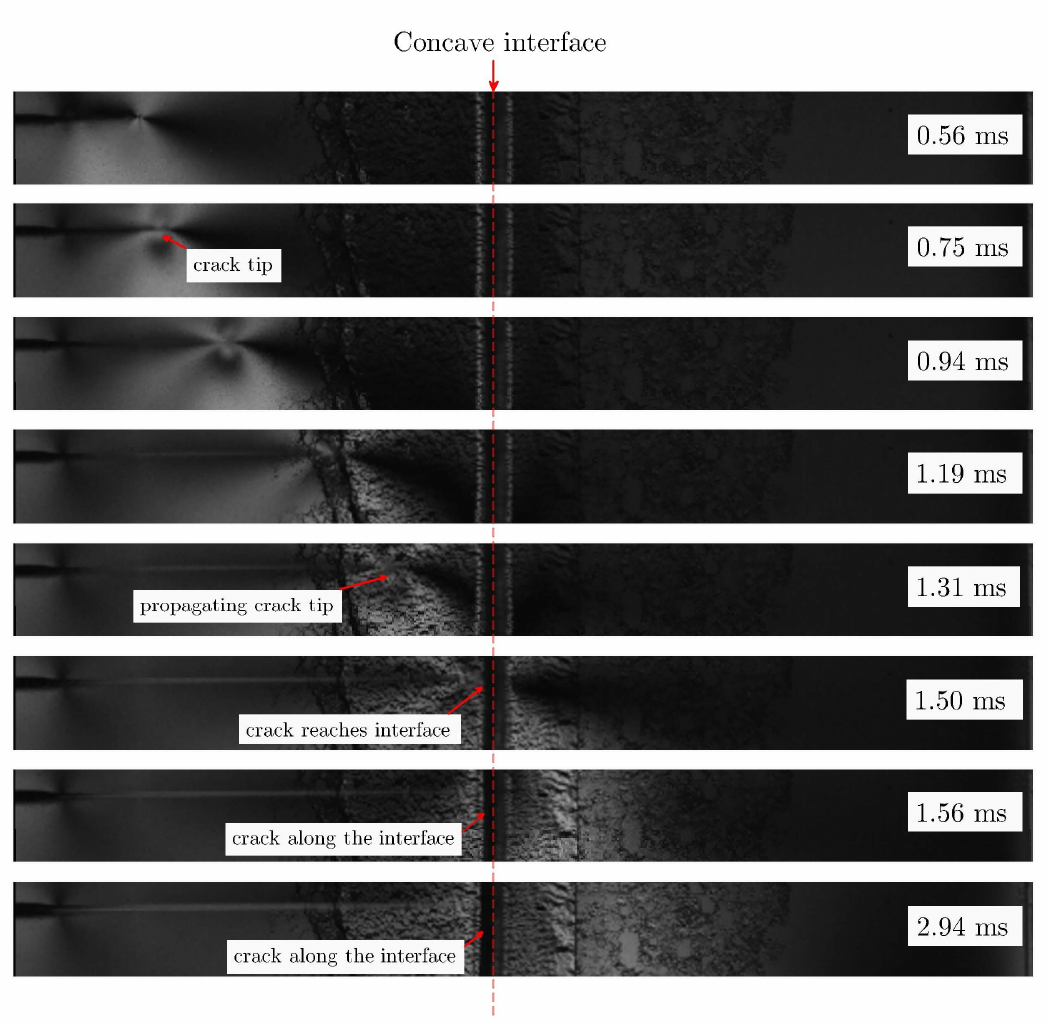}
     \caption{Series of dynamic photoelastic images of crack propagation for a 45 mm radius weak concave interface.}
         \label{fig:detalleconcave}
 \end{figure}

\reffig{fig:detalleconvex} and \reffig{fig:detalleconcave} depict a set of detailed images with the sequence of the initial propagation of the mother crack. The set of images was taken with the optical equipment at 160,000 fps. The red vertical line at the center of each image has been drawn to visually identify the interface position. 

In the convex-interface configuration, \reffig{fig:detalleconvex}, the isochromatic lines are continuous through the interface up to $1.25$ ms, just before the crack reaches the curved interface at $1.44$ ms. This continuity is an evidence of the fact that the interface is perfectly bonded. After that, crack is deflected along the interface and the specimen is separated into two parts ($t= 2.44$ ms).

In the concave-interface configuration, \reffig{fig:detalleconcave}, the propagation of the mother crack is shown for $t= 0.56$ ms, $t= 0.75$ ms, $0.94$ ms, $1.19$ ms and $1.31$ ms, observing a continuity of the isochromatic lines again through the interface. At $t=1.50$ ms, the crack is deflected at the interface. Afterwards, the crack follows the path imposed by the weak interface ($t=1.56$ ms and $2.94$ ms).

\begin{figure}[ht!]
  \centering
     \includegraphics[width=1\columnwidth]{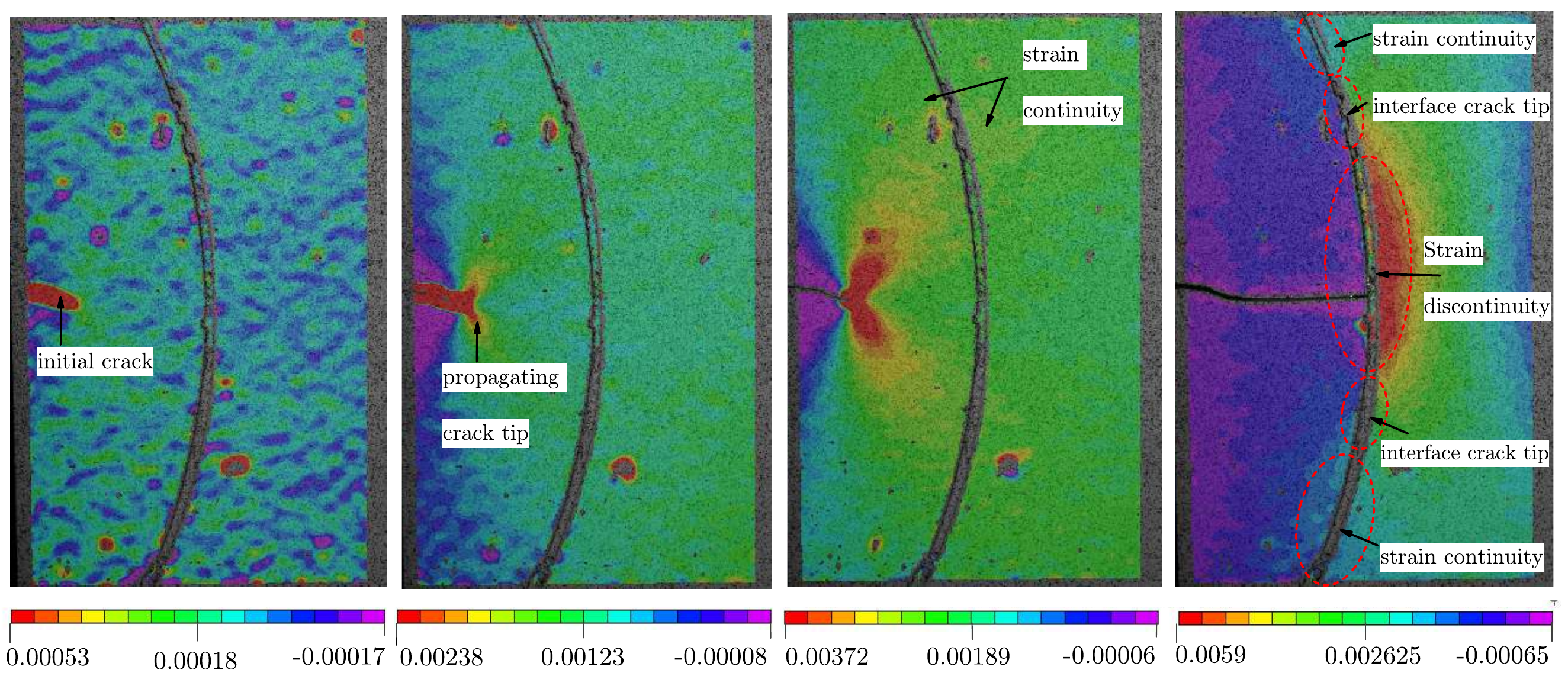}
     \caption{2D digital image correlation representing vertical strain during crack propagation. The discontinuity of strain allows identifying the crack tip positions of the deflected cracks.}
         \label{fig:videocorrelacióndef}
 \end{figure}

To assess the  interface debonding, the digital image correlation (DIC) technique was also employed. A random speckle pattern was sprayed over the specimen surface, and the corresponding  images were acquired before and after the fracture event. \reffig{fig:videocorrelacióndef} shows the map of the vertical strain for a concave-interface specimen. It should be emphasized that during the propagation of the mother crack there is a continuity of strain across the interface, which indicates perfect bonding. Thereby, the  specimen behaved as a single homogeneous material until the crack impinged on the interface and deflected along  it. After that, it is possible to observe where a discontinuity in the vertical strain field took place, which allows   detecting the crack tip positions of the deflected cracks.

\begin{figure}[ht!]
  \centering
     \includegraphics[width=0.7\columnwidth]{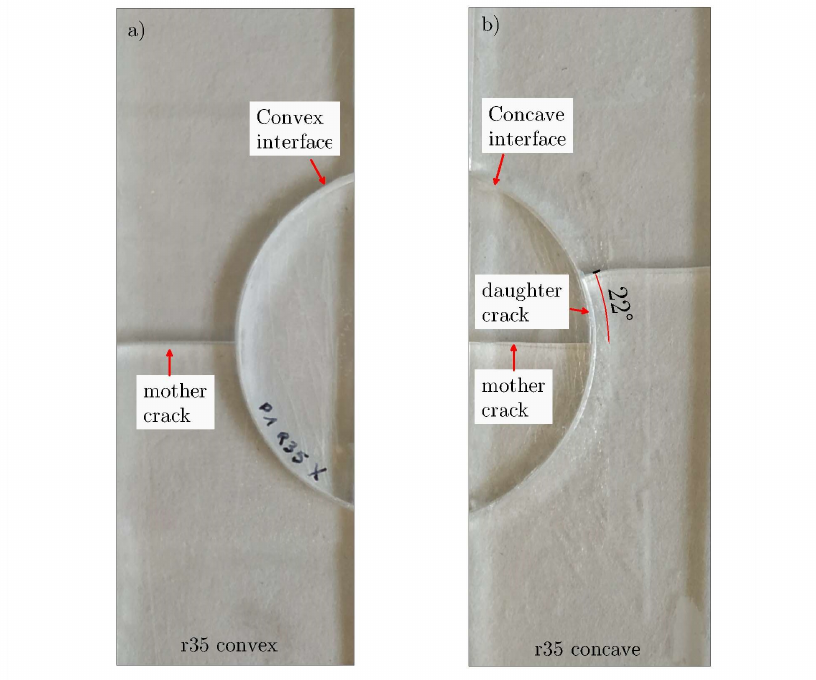}
     \caption{Photographs of fractured specimens, showing crack path in a) convex  and b) concave interface configurations.}
         \label{fig:crackpath}
 \end{figure}

Finally, a post-mortem analysis of the specimens was carried out, studying the crack path in detail for each configuration under analysis. In particular, it is interesting to verify how the crack trajectory was altered by the effect of interface geometry. Typical pictures of two fractured samples, one for each configuration (convex or concave interface), are shown in \reffig{fig:crackpath}. In both cases, it is evident that the crack propagated similarly until it reached the interface and then it branched into two daughter cracks travelling along the interface. At a certain point, crack paths differed from each other. For the concave interface configuration, one daughter crack kinked out of the interface into the adjacent part. However, such  branching was not observed for convex interfaces. The extent of the interface crack growth, given by the polar angle where the crack  kinked out of the interface towards the other part of specimen, varied with the interface curvature radius, as detailed in \reftab{tab:summmary}.

\reffig{fig:crackpath}a shows the crack path for the convex interface configuration.  A straight crack path is observed  until the crack impinges on the weak interface, where it deflected and propagated trapped at  the interface. After the corresponding tests, it was examined that the interface in these configurations was totally debonded, without any kinking out of the interface. \reffig{fig:crackpath}b shows the crack path for the concave interface configuration, where the dynamically growing straight mother crack deflected at the interface, where the daughter interface cracks were trapped for a while, eventually one of them kinked out of the interface   penetrating into the adjacent bulk as a locally mixed-mode daughter crack, as shown in \reffig{fig:crackpath}. 

\begin{table}
 
\begin{center}
 \resizebox{0.99\textwidth}{!}{
 \begin{tabular}{|c|c|c|c|c|c|c|c|c|c|c|}
\hline
              & \multicolumn{10}{c|}{CONVEX}                                                                                       \\ \hline
Curvature radius        & 15 mm    & \multicolumn{2}{c|}{25 mm} & 35 mm    & 45 mm    & \multicolumn{2}{c|}{55 mm} & 65 mm    & \multicolumn{2}{c|}{75 mm} \\ \hline
Specimen         & S1     & S1          & S2         & S1     & S1     & S1          & S2         & S1     & S1       & S2            \\ \hline
Behaviour & unstably & unstably      & unstably     & unstably & unstably & unstably      & unstably     & unstably & unstably   & penetration   \\ \hline 

                           & \multicolumn{10}{c|}{CONCAVE}                                                                                                                                                         \\ \hline
Curvature radius                                & 15 mm                   & \multicolumn{2}{c|}{25 mm}                      & 35 mm                      & 45 mm                      & \multicolumn{2}{c|}{55 mm}                     & 65 mm                      & \multicolumn{2}{c|}{75 mm}                   \\ \hline
Specimen                                 & S1                    & \multicolumn{2}{c|}{S1}                       & S1                       & S1                       & \multicolumn{2}{c|}{S1}                       & S1                       & \multicolumn{2}{c|}{S1}                    \\ \hline
Behaviour                         &                       & \multicolumn{2}{c|}{arrest}                & arrest                & arrest            & \multicolumn{2}{c|}{arrest}               & arrest                &      \multicolumn{2}{c|}{penetration}             \\ \hline
\multicolumn{1}{|l|}{Kinking angle} & \multicolumn{1}{l|}{} & \multicolumn{2}{l|}{$25^\circ$} & \multicolumn{1}{l|}{$22^\circ$} & \multicolumn{1}{l|}{$16^\circ$} & \multicolumn{2}{l|}{$15^\circ$} & \multicolumn{1}{l|}{$18^\circ$} & \multicolumn{2}{l|}{} \\ \hline
\end{tabular}
}
\end{center}
\caption{Summary of the behaviours found in each specimen. Penetration: crack grows straight across the interface. Arrest: crack deflects at the interface and it is arrested at some point of the interface. Unstably: crack  deflects at the interface and grows unstably along the whole interface. Kinking angle: polar angle measured along the interface between the  impinging  and  kinking points, see \reffig{fig:crackpath}.}
\label{tab:summmary}
\end{table}

\reftab{tab:summmary} presents a summary of the behaviour observed in each specimen. It is interesting to notice  that the crack was deviated for all the specimens except for those with the largest radii which failed due to crack penetration, for both concave and convex interface configurations. Thus, it appears  that even a small perturbation of the interface planarity (flatness)  is enough to lead to symmetric crack deflection at the interface as the preferential path. Focusing on the specimens where deflection occurred, crack growth was arrested at some point of the interface in the case of   concave interface configurations. In contrast, for  specimens with convex interfaces, complete debonding of the whole interface took place in an unstable manner without any crack arrest.

From the results presented in this section, it can be concluded that the curved shape of the interface dictated the crack path selection and the fracture behavior. This fact clearly shows that the introduction of a curved interface within the given specimen geometry can deflect a mother crack and also arrest it. 
 
\section{Computational analysis} \label{sec-CurInt1-CompAna}
The main focus of this section is to provide a possible interpretation based on LEFM concepts of the previous experimental results presented in Section 2. To this aim, we use a computational model based on the finite element method.
To double check the present computational results, in addition to a mesh convergence study (not presented here for the sake of brevity),  these results are compared, where possible, to the predictions obtained by using the well-known empirical formulas available in fracture mechanics handbooks. Moreover,  some limit predictions obtained by a  semi-analytical method   available in the literature are compared with the present computational results.

\subsection{Numerical model} \label{sec-CurInt1-NumModel}
\begin{figure}[ht!]
  \centering
     \includegraphics[width=0.8\columnwidth]{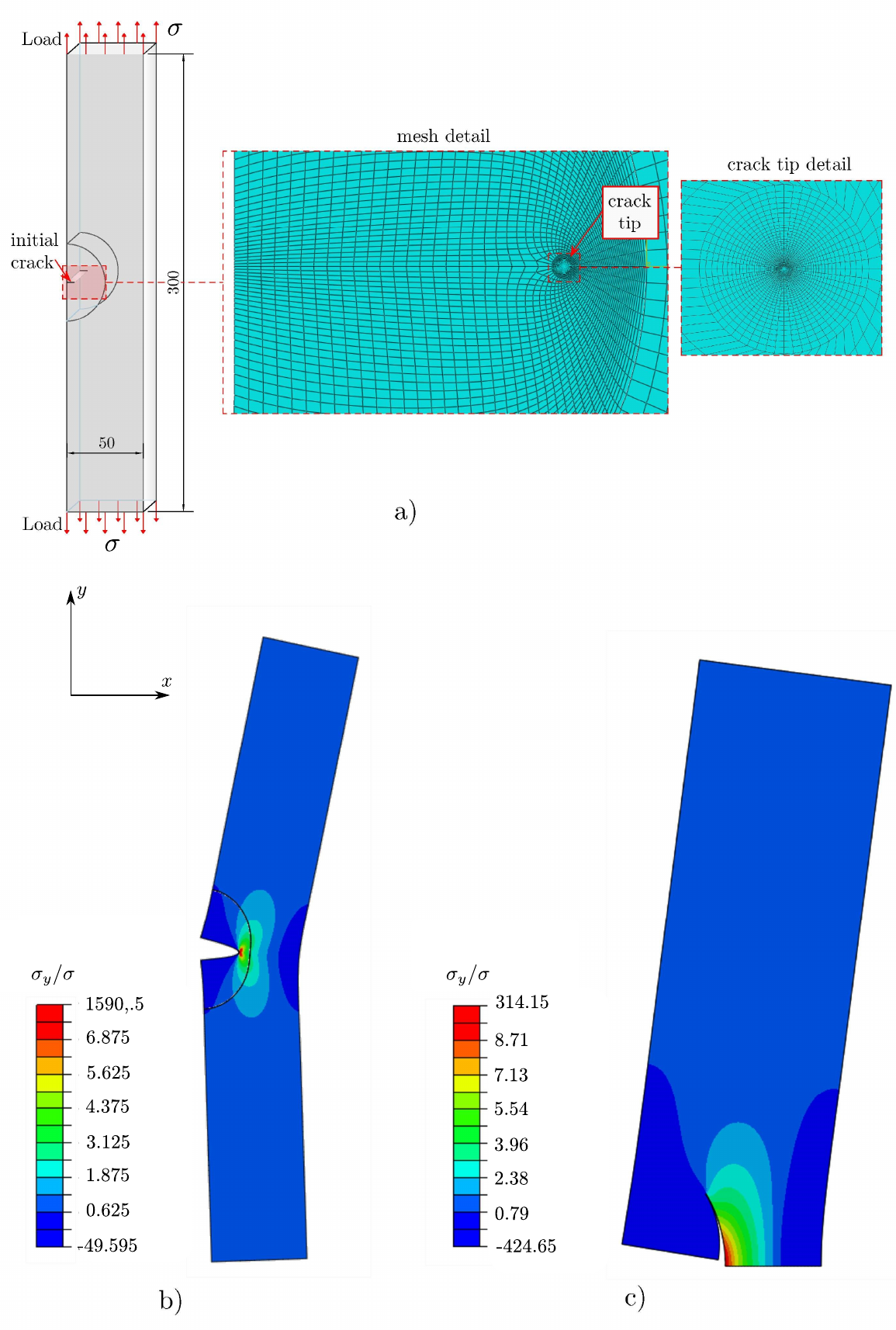}
     \caption{Details of the computational model a) Schematic, b) Results for a mother-crack model c) Results for a daughter-crack model.}
         \label{mesh}
 \end{figure}

The proposed model was based on the Finite Element Method and generated using the commercial FE package \texttt{ABAQUS}. For this purpose, two types of models were developed as shown in \reffig{mesh}: \textit{(i)} for mother-crack path and \textit{(ii)} for daughter-crack path. Plane stress assumptions state was adopted in both models. The geometry definition of the considered specimens   was generated  according to the dimension and shapes defined in the experimental work. The material properties  considered henceforth were: Young's Modulus $3.79$ GPa and Poisson's ratio $0.37$. For the discretization of both models,  finite element meshes constituted by eight node plane stress quadratic elements (CPS8) were used, using the quarter-point isoparametric finite elements around the crack tip. The fracture mechanics parameters characterising  crack propagation, i.e. Stress Intensity Factors (SIFs) and Energy Release Rate components (ERRs), were computed for each given crack tip position by the \textit{J-Integral} that is a built-in function of \texttt{ABAQUS}. A   uniform distributed tensile loading   was applied at the outer extremes of the specimens.   

The mother-crack model was generated by considering  2 \textit{Parts} separated by the interface, where the interaction between the two parts was modeled using \textit{Tie} option, whereas the daughter-crack model was generated with only 1 \textit{Part}. 

In the first model, a partition of the geometry along the horizontal central line was used to model the mother-crack path. The elements on both sides of this discontinuity were separated from each other, i.e. they  shared nodes except for  the crack faces, where separated nodes on elements were defined.
%
In the second model, we simplified the geometry by considering only one half of the specimen, as is depicted in \reffig{mesh}. Then, we propagated the crack along the interface through  untying  the corresponding FE nodes, similarly as done in the first model. 

\subsection{Numerical results} \label{sec-CurInt1-NumRes}
\begin{figure}[ht!]
  \centering
     \includegraphics[width=1\columnwidth]{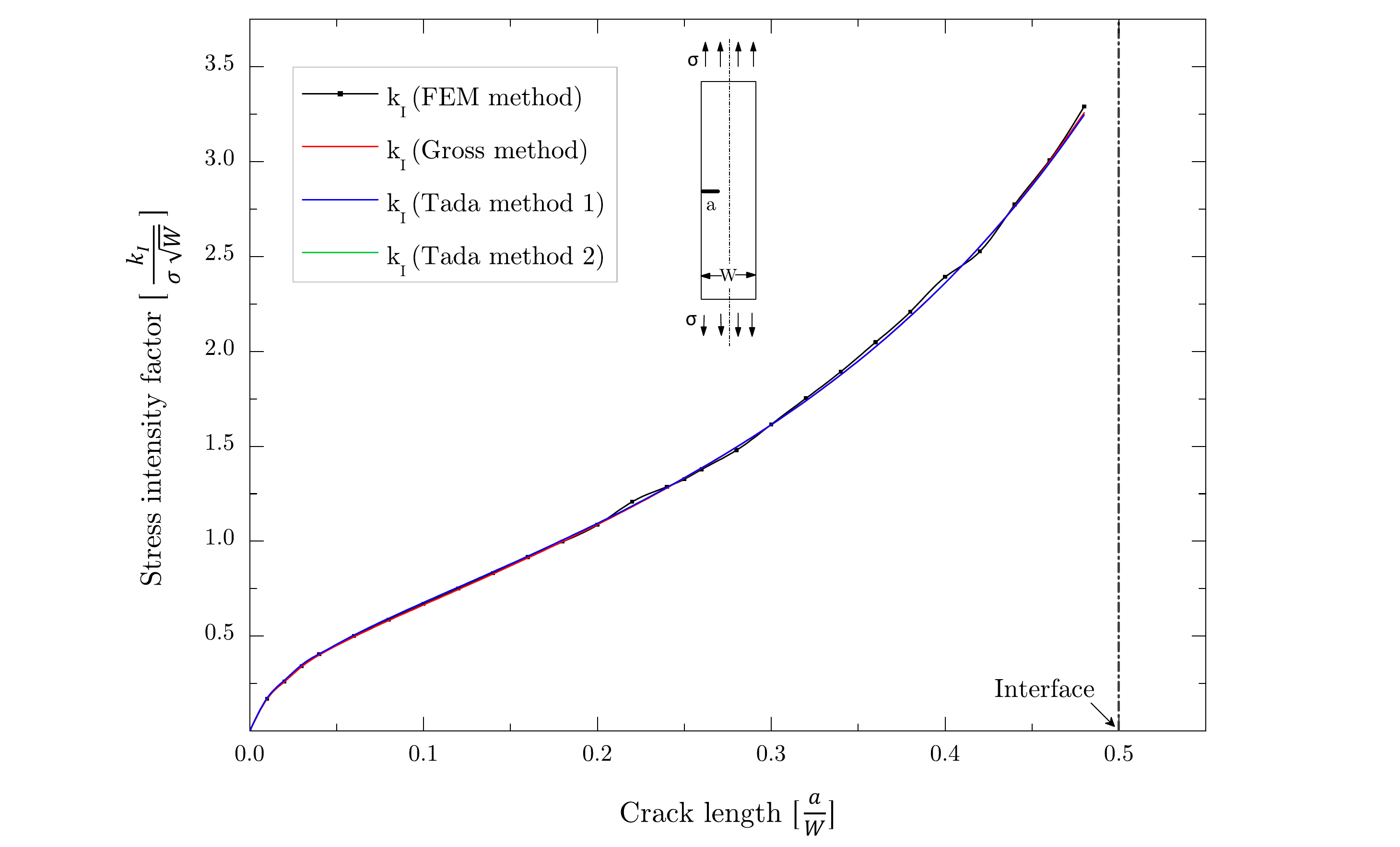}
     \caption{Mode I SIF as a function of the crack length when the mother crack grows straight. Comparison between the results by the computational model and the approximations of the single edge notch test specimen taken from handbooks.}
         \label{Intervalorecton}
 \end{figure}

The evolution of the SIF $k_I$, for the straight propagation of  the mother crack in mode I,    is shown in \reffig{Intervalorecton}. For the sake of accuracy check, the present computational results were compared with the classical approximations by \cite{gross1964,tada-handbook}, see~\refsec{app-sen} for details. This comparison   shows an excellent agreement between the results obtained by both the computational model and the existing benchmark approximations. 

In view of the results shown in \reffig{Intervalorecton}, as expected, the SIF $k_I$ progressively increases with respect to the crack length, which is a necessary condition for an unstable crack growth in a homogeneous material, i.e. without considering any  material interfaces.

\begin{figure}[ht!]
 \centering
     \includegraphics[width=1\columnwidth]{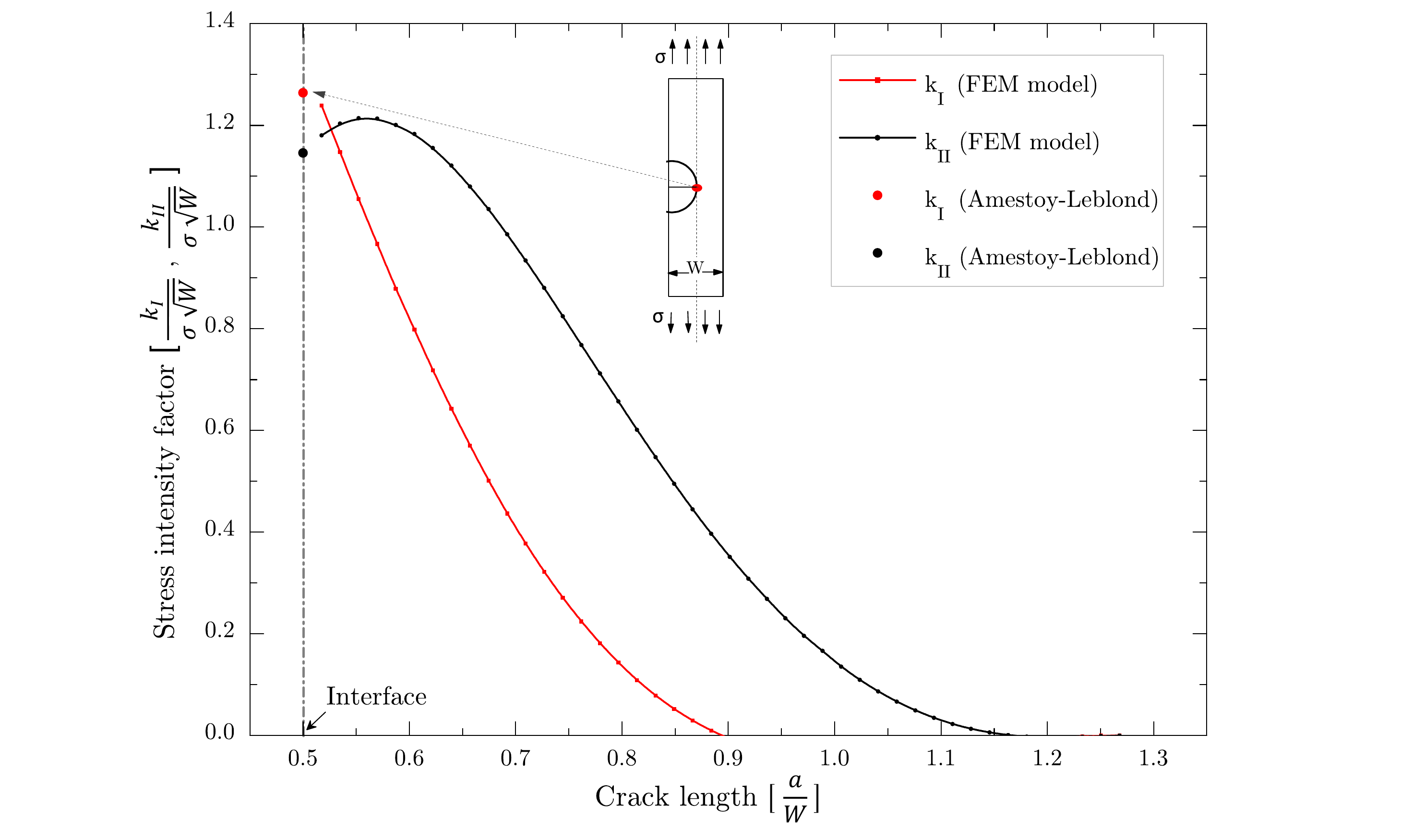}
     \caption{SIFs as a function of the crack length when the daughter cracks grow along the interface for the concave interface   with 25 mm radius.}
         \label{fig:curvo}
 \end{figure}

When the mother crack deflects at the interface, the SIFs evolution changes drastically. \reffig{fig:curvo} shows the SIFs as a function of the crack length when the daughter (interface) cracks propagate along the concave curved path with 25 mm curvature radius. The SIFs computed are   compared with their limit values for  the   daughter crack of vanishing-length kinked by $90^\degree$ angle  obtained  from the semianalytical solution deduced in  \cite{leblond1989}, see the  expressions reported in \ref{app-leblond}. 
Noteworthy, it was recently shown in~\cite{katzav2007} (see Figure 2 therein), that the limit values of SIFs  for the vanishing-length daughter cracks are quite close for   the symmetrically branched crack and for the kinked crack by angle $90^\degree$. 
As can be observed in \reffig{fig:curvo} the computational results are fully consistent with  the semianalytical estimates.

It should be mentioned that for large interface cracks, in the case of concave interface, some overlapping of crack faces may appear, indicated by negative values of SIF $k_I$ (not shown explicitly in \reffig{fig:curvo}), which would predict contact between crack faces, not considered in the present computational model. Nevertheless, this  will   have no influence on the computational-model predictions, due to the fact that the interface cracks  that  typically kinking out towards the other specimen part would   happen    for a shorter interface crack length. This statement is based on the  experimental observations obtained from our tests.

The observed relevant reduction of the SIFs by increasing the crack length implies a stable crack growth. Moreover, the mode II tends to become the  predominant fracture mode. In engineering materials, mode-II crack propagation criteria generally involve higher fracture toughness in comparison to mode-I. Therefore, the above results likely imply higher load-carrying capacities of the concave-interface configurations. 

\begin{figure}[ht!]
  \centering
     \includegraphics[width=1\columnwidth]{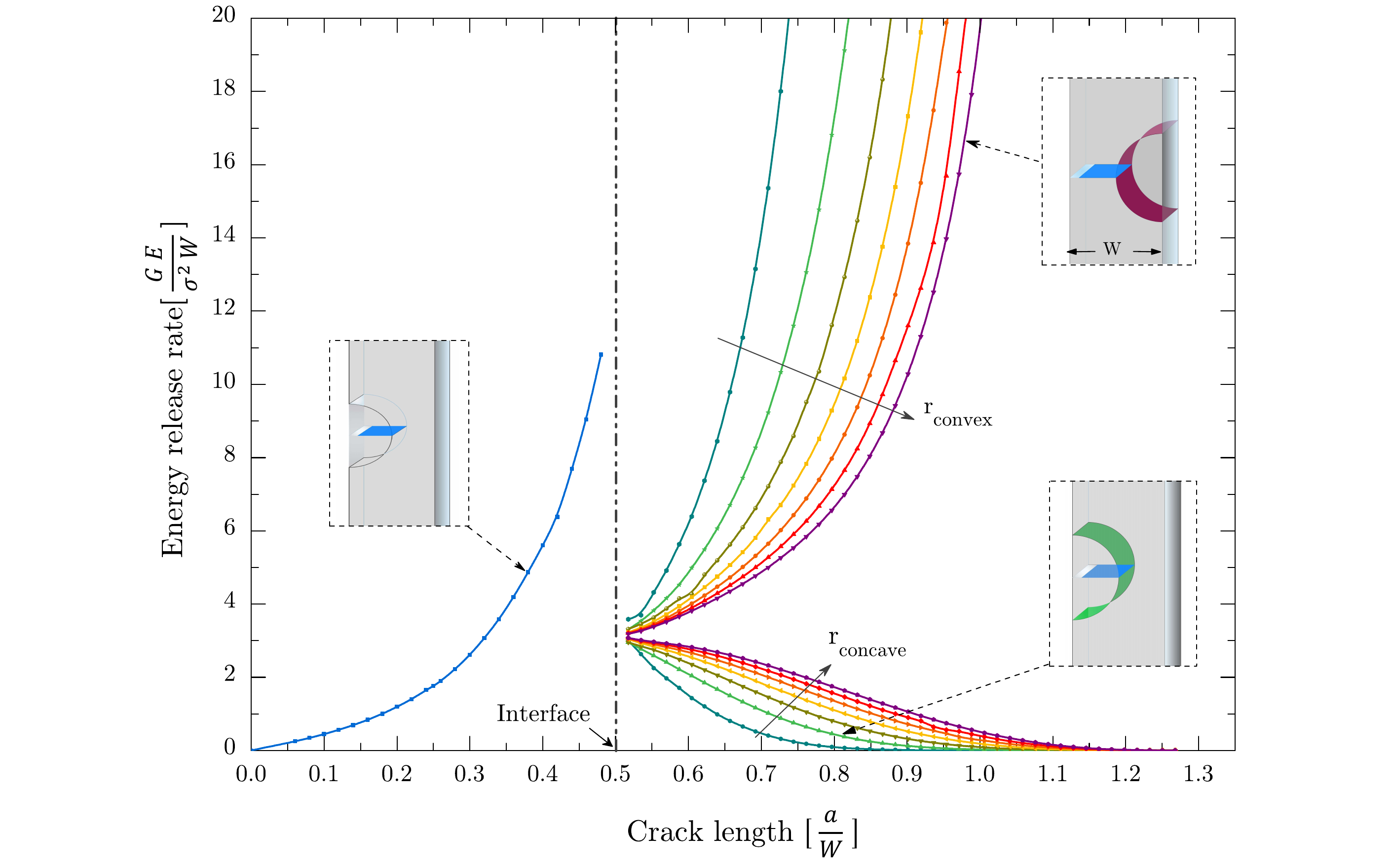}
     \caption{ERR as a function of the crack length, for different potential crack paths.}
         \label{fig:intervalorecto+curvo}
 \end{figure}

In order to evaluate the influence of the curvature radii of the weak interface, the ERR was computed for the different configurations, see \reffig{fig:intervalorecto+curvo}. For the sake of comparison, the ERR for the mother crack growing straight up to the interface is also reported. Note that the represented crack length  corresponds to the sum of the length of the straight crack (mother crack) plus the length of the curved one (one daughter crack). 

The results in \reffig{fig:intervalorecto+curvo} quantitatively clarify the role of the convexity of the weak interface. For convex interfaces, the ERR is strongly increasing, predicting an unstable crack growth. On the other hand, for concave interfaces, the estimations with regard to the ERR features a decreasing evolution, this fact being  related to a stable crack growth. The curvature radius has also a quantitative influence on the ERR behaviour. The general tendency is that a small curvature radius amplifies the effect of the convexity or  concavity of the weak interface. Thus, it can be stated that the differences with respect to the   ERRs estimations  for convex and concave interfaces between two configurations  are larger for smaller  curvature radius than those corresponding to  larger curvature radii.

\section{Discussion and interpretation of results} \label{sec-CurInt1-Disc}
The experimental results can be interpreted thanks to the fracture mechanics data extracted from the computational models. 

The propagation of the mother crack up to meeting the interface took place under mode I, and, as usual,  it occurred  when the corresponding ERR reached the fracture toughness of the bulk. Since the ERR increased with the crack length, the mother crack growth followed an unstable process. When the crack met the interface, its crack path was dictated by the mismatch in the fracture toughness of the bulk and of the interface, that can be assessed by the ratio $G_c^{(i)}/G_c^{(b)}$, as for a straight interface, plus also by the role of the interface curvature, which is an aspect not included in previous theoretical LEFM formulations.

For very tough interfaces, it is likely that the mother crack simply penetrates across the interface in a straight manner, because the ERR is higher if the crack goes straight than if the crack deflects at the interface. However, the experiments revealed  that the mother crack deflected at the interface, see \reftab{tab:summmary}, which is a behaviour typical for a weak interface. From a quantitative standpoint, the debonding event along the interface becomes the actual crack path if the ratio between the energy release rate for the propagation in the bulk and its fracture toughness, $G^{(b)}/G_c^{(b)}$, is lower than the ratio between the energy release rate for double deflection and the interface fracture toughness, $G^{(i)}/G_c^{(i)}$. From the present experimental results, single deflection was not observed.

Afterwards, along the deviated crack path, further competition between debonding and penetration into the second material layer takes place, depending on the interface curvature. In particular, for convex interfaces, the crack growth of the daughter debonding cracks is unstable since the corresponding ERR is a monotonically increasing with the crack length. On the other hand, for concave interfaces, the ERR is progressively decreasing with the progress of crack propagation. At a certain point, crack arrest or penetration can become favourable over further debonding.  

\begin{figure}[ht!]
  \centering
     \includegraphics[width=1\columnwidth]{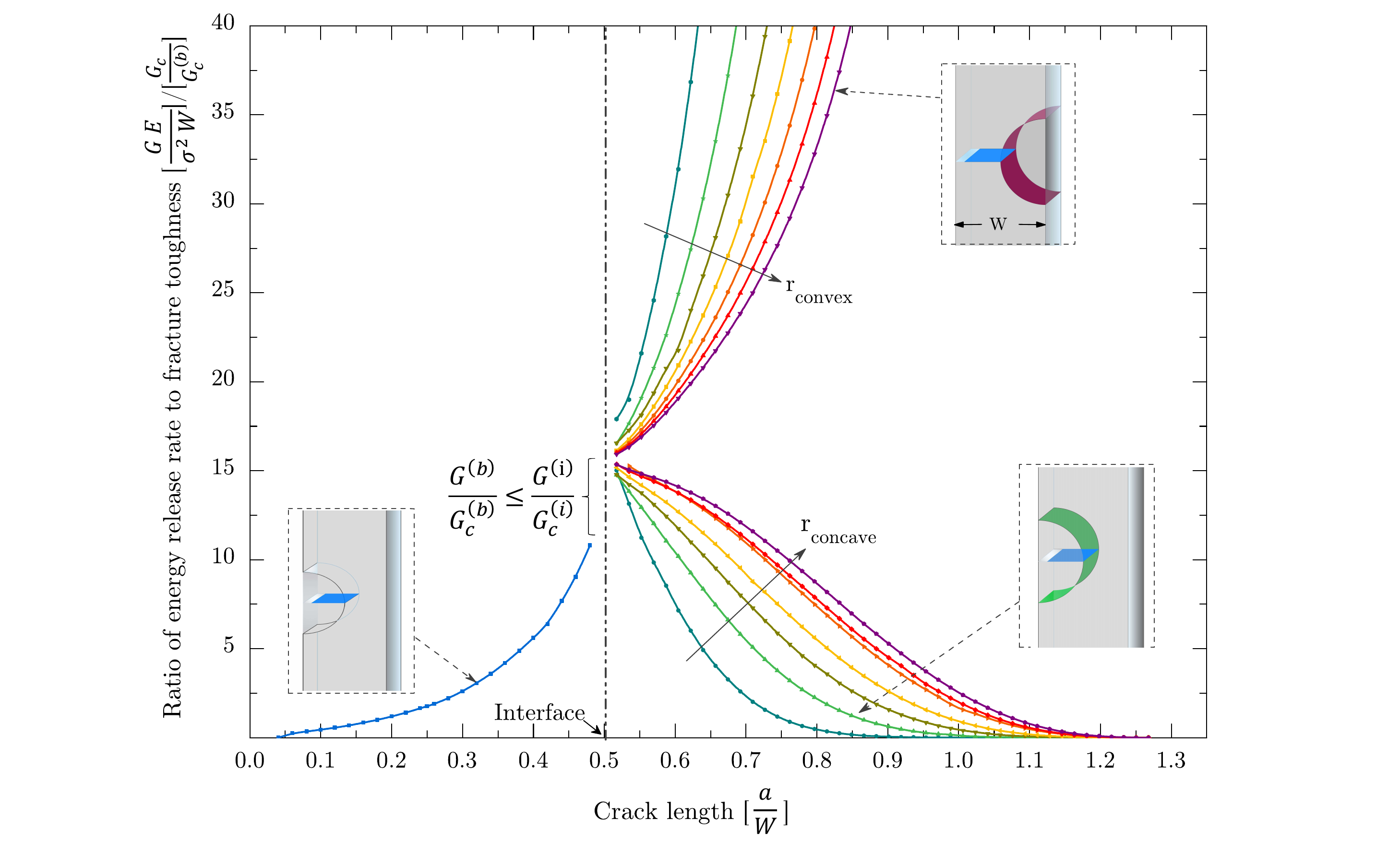}
     \caption{Ratio between the ERRs and $G_{c}/G_c^{(b)}$, where $G_c$ is given by $G_c^{(b)}$ for the mother crack and it corresponds to $G_c^{(i)}$ for the interface.}
         \label{fig:GX5}
 \end{figure}

Although it is not an easy task to characterize the interface toughness from experiments, its qualitative identification can be achieved based on the above concepts by re-scaling the ERR for the mother crack and for the daughter cracks by the factor 
$G_{c}/G_c^{(b)}$, where $G_c$ is given by $G_c^{(b)}$ for the mother crack and it corresponds to $G_c^{(i)}$ for the interface. A possible re-scaled plot that would be consistent with the experimental evidences is shown in \reffig{fig:GX5}  by considering $G_{c}^{(i)}/G_c^{(b)}=1/5$, i.e., for an interface five times weaker than the bulk.

According to \reffig{fig:GX5}, the radius of curvature of the interface  had also an important influence on the crack arrest in concave interfaces. Smaller curvature radii imply lower $G^{(i)}$, thus smaller radii favour the occurrence of crack arrest. This protective phenomenon delays kinking out of the interface towards the bulk, and subsequently the final failure of the specimen. 

\section{Concluding remarks} \label{sec-CurInt1-Conc}
The experimental evidences presented here originally investigated the phenomenon of crack growth in brittle materials with the presence of weak curved interfaces. The current results showed that it is possible to stop/deviate the unstable crack advancement by means of introducing a concave weak interfaces with a sufficiently small radius of curvature. In contrast to alternative strategies for the design of crack arresters, the current concept does not imply adding mass to the structure or holes to deviate the crack paths, which is a remarkable advantage in certain industries where minimization of structural weight is a priority and where the overall structural integrity is a major concern. Moreover, if the strategy is well undertaken, it does not imply necessarily either a reduction of stiffness or the introduction of stress concentrations.

The experimental results were interpreted with the help of FEM computational model of the system incorporating LEFM capabilities. The energy release rate (ERR) extracted from the computations satisfactorily  agreed with well established crack growth criteria at interfaces \citep{he1989} and with respect semi-analytical formulations \citep{leblond1989}. Consequently, these models  provided a suitable methodology to interpret the experimental trends in terms of crack path and its stability, and also to qualitatively identify the interface fracture toughness. 

Both experimental results and computational predictions agree on the fact that the concave weak interfaces are able to arrest the unstable crack growth. The final failure of the whole specimen can occur for a load much higher than that required for the propagation of the initial mother crack, thus significantly improving the damage tolerance of the specimen.  

The results presented here, in particular in what concerns the computational analysis, are illustrating the crack arrest strategy proposed. Further quantitative analyses are indeed necessary in order to investigate other material configurations depending on the interface fracture toughness and also bi-material configurations, being a matter beyond the scope of the current study. From the experimental point of view, the proposed strategy based on material testing combined with photo-elasticity and digital image correlation acquired in real time with a fast digital camera has been found suitable for the present study where the loading rate was sufficiently low. Additional improvements can be requested to test the specimens also in a full dynamic scenario, up to very high strain rates, which is a topic that would deserve a specific attention. 

\section*{Acknowledgements}

The authors would like to express their gratitude to Professor Davide Bigoni (Universita di Trento, Trento, Italy) for stimulating this research. 
This investigation was supported by the Spanish Ministry of Economy and Competitiveness and European Regional Development Fund (Project MAT2015-71036-P, UNSE 15-CE-3581), the Junta de Andaluc\'{i}a and the European Social Fund (Programa de Garant\'ia Juvenil 2017), and   the Spanish Ministry of Science, Innovation and Universities and European Regional Development Fund (Project PGC2018-099197-B-I00).
 MP would like to acknowledge the support from the Italian Ministry of Education, University and Research to the Research Project of Relevant National Interest(PRIN-2017) XFAST-SIMS: Extra-fast and accurate simulation of complex structural systems. 

\appendix

\section{Expressions used for verification purposes of the ERR}

\subsection{Single edge notch} \label{app-sen}

For single-edge-notched tensile specimen with width $b$ with a crack with length $a$, open to the left free boundary of the plate, and with uniform tension $\sigma$, applied across the upper and lower boundaries of the plate (boundaries parallel to the crack), the value of the Stress Intensity Factor can be approximated by:
\begin{equation} \label{eu_eqn}
k_I= \sigma^2+\sqrt{ \pi a } F(a/b)
\end{equation}
where F(a/b) is:
\begin{itemize}
\item  \cite{gross1964}:
\begin{equation} \label{eu_eqn2}
F(a/b)= 1.122 -0.231(a/b) + 10.550(a/b)^2-21.710(a/b)^3 +30.382(a/b)^4
\end{equation}
with accuracy of 0.5\% for $a/b \leq 0.6$.

\item \cite{tada-handbook}:
\begin{equation}
F(a/b)= 0.265(1-a/b)^4 + \frac{0.857+0.265 a/b}{(1-a/b)^{3/2}}
\end{equation}
with accuracy better than $1\%$ for $a/b < 0.2$, and $0.5\%$ for $a/b \geq 0.2$.

\item \cite{tada-handbook}
\begin{equation} \label{eu_eqn3}
F(a/b)= \sqrt{\frac{2b}{\pi a} \tan \frac{\pi a}{2b}} \frac{0.752+2.02(a/b)+0.37(1-\sin \frac{\pi a}{2b})^3}{cos \frac{\pi a}{2b}}
\end{equation}
with accuracy of $0.5\%$ for any $a/b$.
\end{itemize}

\subsection{Stress intensity factors just after the kink} \label{app-leblond}
According to \cite{leblond1989,amestoy1992}, in plane situation, the stress intensity factor of a kinked crack just after the kink $k_p^*$, i.e. in the limit when the kink length tends to zero, can be expressed as a function of the stress intensity factor just before the kink, $K_q$, and the kink angle $m \pi$ (expressed in radians):
\begin{equation} \label{eu_eqn4}
k_p^* = F_{pq}(m)  K_q
\end{equation}
where:
\begin{dmath} 
F_{11} (m) = 1 - \frac{3 \pi^2}{8}  m^2 + (\pi^2 - \frac{5 \pi^4} {128} ) m^4 + (\frac{\pi^2} {9} - \frac{11 \pi^4} {72} + \frac{119 \pi^6} {15360}) m^6 + 5.07790 m^8 - 2.88312 m^{10} - 0.0925 m^{12} + 2.996 m^{14} - 4.059 m^{16} + 1.63 m^{18} + 4.1 m^{20}
\end{dmath}
\begin{dmath} 
F_{22} (m) = 1 - (4 + \frac{3 \pi^2} {8}) m^2 + (\frac{8}{3} + \frac{29 \pi^2}{18} - \frac{5 \pi^4}{128}) m^4 + (-\frac{32}{15} - \frac{4 \pi^2}{9} - \frac{1159 \pi^4}{7200} + \frac{119 \pi^6}{15360}) m^6 + 10.58254  m^8 - 4.78511 m^{10} - 1.8804 m^{12} + 7.280 m^{14} - 7.591 m^{16} + 0.25 m^{18} + 12.5 m^{20}
\end{dmath}
\begin{dmath}  
F_{12}(m) = -\frac{3 \pi}{2} m + (\frac{10 \pi}{3} + \frac{\pi^3}{16}) m^3 + (-2 \pi - \frac{133 \pi^3}{180} + \frac{59 \pi^5}{1280}) m^5 + 12.313906 m^7 - 7.32433 m^9 + 1.5793 m^{11} + 4.0216 m^{13} - 6.915 m^{15} + 4.21 m^{17} + 4.56 m^{19}
\end{dmath}
\begin{dmath} 
F_{21}(m) = \frac{\pi}{2} m - (\frac{4 \pi}{3} +\frac{\pi^3}{48}) m^3 + (-\frac{2 \pi}{3} + \frac{13 \pi^3}{30} - \frac{59 \pi^5}{3840}) m^5 - 6.176023 m^7 + 4.44112 m^9 - 1.5340 m^{11} - 2.0700 m^{13} + 4.684 m^{15} - 3.95 m^{17} - 1.32 m^{19}
\end{dmath}

\section*{References}
\bibliographystyle{unsrt}
\bibliography{references}

\end{document}